\documentclass[12pt]{article}
\usepackage[numbers]{natbib}
\usepackage[english]{babel}
\usepackage{amsmath}
\usepackage{amssymb}
\usepackage{amsthm}
\usepackage{mathtools}
\usepackage{authblk}
\usepackage{hyperref}
\usepackage{bm}
\usepackage{tikz}
\usetikzlibrary{arrows,positioning,shapes.geometric,decorations.markings}
\tikzstyle{squarenode}=[rectangle,draw]
\tikzstyle{littlenode}=[circle,draw,minimum size=0.5cm,font=\small]
\tikzstyle{bigellipse}=[ellipse,draw,x radius=10cm, y radius=5cm]
\tikzset{negate/.style={
            decoration={markings,
            mark= at position 0.5 with {
                  \node[transform shape] (tempnode) {$\Big\Vert$};
                  }
              },
              postaction={decorate}
}
}
\usepackage{tabularx}
\usepackage{multirow}
\usepackage{booktabs}
\usepackage{subfig}
\usepackage{soul}
\newcommand{\nind}{\mbox{$\perp\!\!\!\perp\!\!\!\!\!\!\setminus$}}
\newcommand{\ind}{\mbox{$\perp\!\!\!\perp$}}
\newcommand{\sst}[1]{\scriptscriptstyle \textup{#1}}

\newcolumntype{Y}{>{\centering\arraybackslash}X}

\def\ind{\perp\!\!\!\perp}
\def\nind{\perp\!\!\!\!\!/\!\!\!\!\!\perp}

\begin{document}

\title{Mediation analysis with case-control sampling: Identification and estimation in the presence of a binary mediator}

\author[1]{Marco Doretti}
\author[2]{Minna Genb{\"a}ck}
\author[3]{Elena Stanghellini}

\affil[1]{Department of Political Science, University of Perugia, 06123 Perugia, Italy}
\affil[2]{School of Economics, Business and Statistics, University of Ume\aa, 90187 Ume\aa, Sweden}
\affil[3]{Department of Economics, University of Perugia, 06123 Perugia, Italy}

\date{}

\maketitle

\begin{abstract}
With reference to a stratified case-control procedure based on a binary variable of primary interest, we derive the expression of the distortion induced by the sampling design on the parameters of the logistic model of a secondary variable. This is particularly relevant when performing mediation analysis (possibly in a causal framework) with stratified case-control data in settings where both the outcome and the mediator are binary. Our identification result opens the way to M-estimation and Maximum Likelihood estimation. We then conduct a simulation study showing the gain in efficiency of the estimators of both the outcome and mediator model parameters w.r. to existing methods, based on weighting. As an illustrative example, we reanalyze a German case-control dataset in order to investigate whether the effect of reduced immunocompetency on listeriosis onset is mediated by the intake of gastric acid suppressors.

\noindent {\bf Keywords}: binary variables, collider node, distortion, odds-ratio, logistic regression,  secondary outcome
\end{abstract}


\newpage

\section{Introduction}\label{sec:intro}
Retrospective sampling schemes are common practice in studies when either the outcome of interest is rare or some covariates are difficult to measure. In this context, it is too expensive or too lengthy to random sample the whole population. Instead, more efficiently, random samples are extracted from a partition of the population according to the outcome and possibly other stratifying factors. When the outcome is binary, this procedure is known as (stratified) case-control sampling design.
 
Though the implications of the retrospective sampling are widely understood when the only interest is the relationship between the covariates and the outcome~\citep{Breslow1996}, less is known when the relationship between some of the covariates is also the object of inference. Typically, this situation arises in mediation analysis, where the aim is to decompose, in a counterfactual framework, the total effect of a treatment on the outcome into a direct and an indirect effect, the second one due to a mediator. A mediator is a variable that is a response of the treatment and in turn influences the outcome~\citep{Pearl2001,VdW2015}. In this situation, two equations are of interest, the first linking the response to the treatment and the mediator, the second linking the mediator to the treatment. If the probability of a unit to be in the sample depends on the outcome (and possibly some other stratifying factors), the relationship between the treatment and the mediator is bound to be distorted and standard mediation methods fail.
 
To tackle this problem, VanderWeele and Vanstelaandt (2010)~\citep{VdWVansteelandt2010} propose to estimate the mediator-treatment equation on the sub-population of controls. Under the rare outcome assumption, this can be considered a random sample from the whole population; see also VanderWeele and Tchetgen Tchetgen (2016)~\citep{VdWTchetgenTchetgen2016}. Another strategy, also mentioned in VanderWeele and Vanstelaandt (2010)~\citep{VdWVansteelandt2010}, is based on weighting, a proposal originally contained in some papers by Mansky and coauthors~\citep{ManskiLerman1977,ManskiMcFadden1981} in the context of \emph{choice-based} samples, which are the analogous of stratified case-control samples within the econometric literature; see also VanderLaan (2008)~\cite{VanderLaan2008}.
 
With reference to a binary mediator and a binary outcome, we here propose a parametric approach to mediation analysis when the sampling scheme depends on the outcome, and possibly other stratifying covariates. We build on results already existing in the literature on parametric mediation analysis for a binary mediator and a binary outcome~\citep{StanghelliniDoretti2019,Dorettietal2022} to derive the explicit expression of the distortion induced by the sampling design. We exploit the knowledge provided by Directed Acyclic Graph (DAGs)~\citep{LauritzenBook}, where an additional binary node is representing the sampling scheme. In this regard, we are in line with Didelez et al. (2010)~\citep{Didelezetal2010}. We then propose two estimating procedures, one based on Maximum Likelihood (ML) and the other one on M-estimation, and compare them with the weighting method.
 
There are a number of additional situations where this investigation can be of interest. The first one concerns informative missingness in logistic regression. It turns out that our work extends the derivations in Wang et al. (2107)~\citep{Wangetal2017} to the continuous treatment context, also proposing two likelihood-based estimating procedures that can be easily implemented with standard statistical software. Another situation of interest arises in case control studies with secondary or additional outcomes~\citep{Richardsonetal2007}, which is particularly common in genetic epidemiology, where information on a secondary phenotype is also of interest~\citep{WangShete2011}. More generally, properly recovering the distortion induced on associations by conditioning on a collider node is fundamental for many other purposes. In this sense, this work extends to a parametric framework the results on odds-ratios of Nguyen et al. (2019)~\citep{Nguyenetal2019}.

The reminder of the paper is organized as follows. In Section~\ref{sec:background}, we set notation and introduce the target regression models, showing that adjustments are needed to identify them in the presence of outcome-dependent sampling schemes. Our identification strategy is presented in Section~\ref{sec:ident}, while the specifics of the two estimation methods deriving from it (M-estimation and ML) are described in Section~\ref{sec:est}. In Section~\ref{sec:appl}, we apply the proposed methodology to data coming from a German case-control study on listeriosis. Our aim is to decompose, in a counterfactual framework, the effect of reduced immunocompetency on listeriosis into a direct and an indirect effect, with the latter mediated by gastric acid suppressor intake. Section~\ref{sec:sims} reports evidence from a simulation study conducted to compare M- and ML estimators to weighting estimators, while some concluding remarks are given in Section~\ref{sec:concl}.


\section{Background}\label{sec:background}
Let $A$ denote a generic treatment/exposure, while $M$ and $Y$ represent the binary mediator and the binary outcome, respectively. Like in other related approaches (see e.g. Didelez et al., 2010~\citep{Didelezetal2010}), outcome-dependent sampling is theoretically represented by a selection indicator variable, $W$. Data are available just on the selected units, i.e. those $W=1$. A straightforward extension, that is not here considered, is when data can be seen as a random sample extracted from the population with $W=1$.

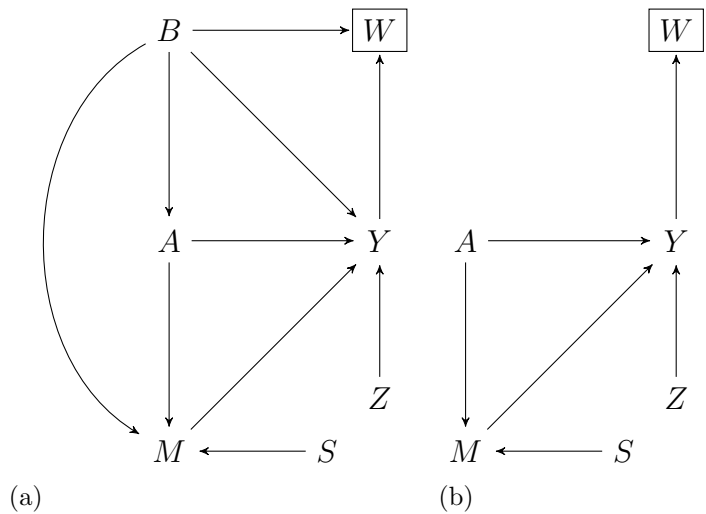
\begin{figure}[tb]
\centering{
\subfloat[][]{\label{fig:medccbasecov}}
\begin{tikzpicture}[scale=0.5,auto,->,>=stealth',shorten >=1pt,node distance=2.8cm] 
\node (X) {$A$};
\node (Y) [right of=X]{$Y$};
\node (M) [below of=X] {$M$};
\node (S) [below of=Y,xshift=-0.7cm] {$S$};
\node (Z) [below of=Y,yshift=0.7cm] {$Z$};
\node[squarenode] (W) [above of=Y] {$W$};
\node (B) [left of=W] {$B$};
\draw[->] (X) --node {} (Y); 
\draw[->] (X) --node {} (M); 
\draw[->] (M) --node {} (Y); 
\draw[->] (Y) --node {} (W);
\draw[->] (B) --node {} (X); 
\draw[->] (B) --node {} (Y); 
\draw[->] (B) --node {} (W);
\draw[->] (B) to [bend right=60] (M);
\draw[->] (S) --node {} (M);
\draw[->] (Z) --node {} (Y);
\end{tikzpicture}\quad
\subfloat[][]{\label{fig:medccbasenocov}}
\begin{tikzpicture}[scale=0.5,auto,->,>=stealth',shorten >=1pt,node distance=2.8cm] 
\node (X) {$A$};
\node (Y) [right of=X]{$Y$};
\node (M) [below of=X] {$M$};
\node (S) [below of=Y,xshift=-0.7cm] {$S$};
\node (Z) [below of=Y,yshift=0.7cm] {$Z$};
\node[squarenode] (W) [above of=Y] {$W$};
\draw[->] (X) --node {} (Y); 
\draw[->] (X) --node {} (M); 
\draw[->] (M) --node {} (Y); 
\draw[->] (Y) --node {} (W);
\draw[->] (S) --node {} (M);
\draw[->] (Z) --node {} (Y);
\end{tikzpicture}
}
\caption{Mediation scheme for the $(A,M,Y)$ triplet and covariates $(S,Z)$ in an unconditional (a) and stratified (b) case-control sampling setting. Note that variables in $B$ may influence any set (i.e. not necessarily all) of the variables $W$, $A$, $M$ and $Y$ , and that variables in $S$ may be in $Z$ and vice versa.}
\label{fig:medccbase}
\end{figure}

The binary variable $W$ is influenced by $Y$ via a known probabilistic mechanism, that can be either marginal or conditional on the strata formed by a set of categorical background covariates, $B$, that possibly affect $A$, $M$ and/or $Y$ as well. When $B$ is empty, this corresponds to the unconditional Case-Control (CC) design. Otherwise, the Stratified Case-Control (SCC) design arises. We also indicate by $S$ and $Z$ the sets of covariates (of any nature) that influence $M$ and/or $Y$, respectively, but not the selection mechanism $W$. Notice that $S$ and $Z$ do not necessarily have to be disjoint sets, i.e., variables in $S$ might also be in $Z$ and vice versa. The overall framework is represented in the two DAGs in Figure~\ref{fig:medccbase}, where the $W$ node is squared in order to pinpoint its nature of selection node.


In what follows, we account for the presence of $B$ and thus refer to the SCC setting, with the CC one being a special case thereof. In order to ease the exposition, without loss of generality we think of $B$, $S$ and $Z$ as of singleton variables, with $B$ taking values in $\{1,\dots,N_b\}$ to represent the $N_b$ different strata formed by background variables. Also, if ambiguities do not occur we occasionally use $P(\cdot\mid k)$ as a shorthand for $P(\cdot\mid K=k)$ . The logistic regression models assumed to hold in the population are
\begin{equation}\label{eq:ypop}
\textup{logit}\{P(Y=1\mid a,b,z,m)\} = \bm{x}_y^\top\bm{\beta}
\end{equation}
and
\begin{equation}\label{eq:mpop}
\textup{logit}\{P(M=1\mid a,b,s)\} = \bm{x}_m^\top\bm{\delta},
\end{equation}
where $\bm{\beta}$ and $\bm{\delta}$ are coefficient vectors, while $\bm{x}_y$ and $\bm{x}_m$ contain the values of the explanatory variables (including the constant term) as well as stratum-specific indicator variables $\mathbb{I}\{B=b\}$ for $b=2,\dots,N_b$ (the usual corner-point parametrization taking the first stratum as reference is adopted). Clearly, such a general formulation allows for the presence, in both models, of first- and higher-order (e.g., quadratic) effects as well as of interaction terms. The impact of SCC sampling can be expressed in the language of DAGs and conditional independence~\citep{Dawid1979} by stating that $Y$ and $M$ are not conditionally independent of $W$ given the respective parent node sets, i.e.,
\[
\begin{split}
Y&\nind W \mid (A,B,M,Z) \\
M&\nind W \mid (A,B,S).
\end{split}
\]
As a consequence,
\[
\begin{split}
\textup{logit}\{P(Y=1\mid a,b,z,m,W=1)\} &\neq \bm{x}_y^\top\bm{\beta} \\
\textup{logit}\{P(M=1\mid a,b,s,W=1)\} &\neq \bm{x}_m^\top\bm{\delta},
\end{split}
\]
which shows that the models postulated in the right-hand sides of~\eqref{eq:ypop} and~\eqref{eq:mpop} are not identified in the sub-population of selected units, and that adjusting approaches are needed. 
However, by inspection of the DAG it is possible to see that 
\begin{equation}\label{eq:ci}
M\ind W \mid (A,B,S,Z,Y).
\end{equation}
\noindent and therefore $Y$ can be seen as a \emph{selection bias breaking} variable, a term introduced elsewhere in the literature~\citep{Genelettietal2009}. This property, together with the explicit expressions of the distortion induced by the sampling design on the parameters of~\eqref{eq:ypop} and~\eqref{eq:mpop}, forms the basis of our proposal. 
Notice that the minimum conditioning set for independence of $M$ and $W$ to hold is $(B,Y)$. However, since we are interested in the parametric formulation of the model for $M$, all variables in the conditioning set of \eqref{eq:ci} are relevant.


\section{Identification}\label{sec:ident}
In order to identify the coefficients of model~\eqref{eq:ypop} from SCC data, it is well-known that simple coefficient adjustments involving the conditional population prevalences $\pi_b=P(Y=1\mid B=b)$ suffice~\citep{PrenticePyke1979,FearsBrown1986,Breslowetal1988}. In detail, the model expression holding in the sub-population of selected units is given by
\begin{equation}\label{eq:y}
\textup{logit}\{P(Y=1\mid a,b,z,m,W=1)\} = \bm{x}_y^\top\bm{\beta}^\star,
\end{equation}
where $\bm{\beta}^\star$ is equal to $\bm{\beta}$ except for modifications concerning the intercept ($\beta_{\sst{INT}}$) and the coefficient sub-vector $\bm{\beta}_b=(\beta_2,\dots,\beta_{N_b})^\top$ related to the $N_b-1$ stratum indicator variables. Specifically, letting $p_b=P(Y=1\mid W=1,B=b)$ be the proportion of cases in the SCC sample for stratum $b$, the replacing elements in $\bm{\beta}^\star$ are
\begin{equation}\label{eq:parcor}
\begin{split}
\beta_{\sst{INT}}^{\star} &= \beta_{\sst{INT}}+\log(k_1) \\
\bm{\beta}_b^{\star} &= \bm{\beta}_b+\log(\bm{k}_{-1}),
\end{split}
\end{equation}
where $\bm{k}_{-1}=(k_2,\dots,k_{N_b})^\top$ and, for $b\in\{1,\dots,N_b\}$,
\[
\begin{split}
k_b &= \frac{P(W=1\mid Y=1,B=b)}{P(W=1\mid Y=0,B=b)} \\
&= \frac{P(Y=1\mid W=1,B=b)}{P(Y=0\mid W=1,B=b)}\cdot\frac{P(Y=0\mid B=b)}{P(Y=1\mid B=b)} = \frac{p_b}{1-p_b}\cdot\frac{1-\pi_b}{\pi_b}.
\end{split}
\]
Each of the correction terms above is known whenever $\pi_b$ is. Indeed, $p_b/(1-p_b)$ is the stratum-specific CC ratio, which is fixed by design. 
We here assume $\pi_b$ ($b=1,\dots,N_b$) to be known from external auxiliary information sources. 

We turn now to the model for $P(M=1\mid a,b,s,z,y, W=1)$. From \eqref{eq:ci}, we know that $P(M=1\mid a,b,s,z,y, W=1)=P(M=1\mid a,b,s,z,y)$.
After some derivations, see Appendix~\ref{app:der45}, it is possible to see that 
\begin{equation}\label{eq:mygennow}
\textup{logit}\{P(M=1\mid a,b,s,z,y)\} = \bm{x}_m^\top\bm{\delta} + o(y,\bm{x}_{y};\bm{\beta}),
\end{equation}
where 
\begin{equation}
o(y,\bm{x}_{y};\bm{\beta})=\log\frac{P(Y=y\mid a,b,z,M=1)}{P(Y=y\mid a,b,z,M=0)}
\end{equation}
is a correction term depending on $y$ and the coefficient vector $\bm{\beta}$. The parametric expression of it is \eqref{eq:proof451} of Appendix~\ref{app:der45}. 



\section{Estimation}\label{sec:est}
Starting from the identification result in~\eqref{eq:y}, inference on $\bm{\beta}^{\star}$ can be conducted from SCC data in a standard ML framework~\citep{PrenticePyke1979,Anderson1972}. In parallel to this coefficient adjusting approach, weighting methods have also been proposed within the econometric literature dealing with \emph{choice-based} samples~\citep{ManskiLerman1977,ManskiMcFadden1981}, which can essentially be thought of as the analogous of SCC samples~\citep{Breslow1996}. These methods consist in fitting the population model~\eqref{eq:ypop} to the SCC sample, assigning cases and controls in each stratum $b$ a weight equal to $\pi_b/p_b$ and $(1-\pi_b)/(1-p_b)$, respectively. The resulting estimators are consistent but, although possibly less prone to model misspecification~\citep{XieManski1989}, they were found to be typically less efficient than the ones obtained with parameter corrections~\citep{KingZeng2001}.

Importantly, the weighting approach was also extended to variables different from the original outcome (sometimes termed secondary or additional outcomes~\citep{Richardsonetal2007}); see for example the more general work by VanderLaan (2008)~\cite{VanderLaan2008}. Such an extension is of particular relevance in a mediation framework, where the population parameters of the mediator model also need to be recovered from SCC data in order to achieve effect decomposition~\citep{VdWVansteelandt2010}. In our setting, this would correspond to fitting the population model~\eqref{eq:mpop} to the SCC sample with the same weighting scheme as above. 

Here, we introduce an alternative framework where estimation of the $\bm{\delta}$ vector from SCC data exploits the identification result in~\eqref{eq:mygennow}. Since $\bm{\beta}$ is also involved in that equation, our framework is designed for estimating the combined parameter vector $\bm{\theta}=(\bm{\beta}^\top,\bm{\delta}^\top)^\top$ altogether. This also fits with the mediation setting, where typically both $\bm{\beta}$ and $\bm{\delta}$ have to be estimated at the same time. Two estimation strategies are discussed: in Section~\ref{subsec:mest} we present an M-estimation approach, while in Section~\ref{subsec:mle} the above mentioned ML framework is extended to our bivariate context.


\subsection{M-estimation}\label{subsec:mest}
In a formal M-estimation setting~\citep{Huber1964}, the M-estimator $\hat{\bm{\theta}}_{\sst{M}} = (\hat{\bm{\beta}}_{\sst{M}}^\top,\hat{\bm{\delta}}_{\sst{M}}^\top)^\top$ is the vector satisfying
\[
\bm{\psi}(\hat{\bm{\theta}}_{\sst{M}}) = \begin{pmatrix} \bm{s}_y(\hat{\bm{\theta}}_{\sst{M}}) \\ \bm{s}_m(\hat{\bm{\theta}}_{\sst{M}}) \end{pmatrix} = \bm{0}_{2d_{\theta}},
\]
where $d_{\theta}$ is the dimension of $\bm{\theta}$ and $\bm{s}_y(\cdot)$ and $\bm{s}_m(\cdot)$ are the observed score functions associated to the models in~\eqref{eq:y} and~\eqref{eq:mygennow}, respectively; see Appendix~\ref{app:mest} for their expressions. In practice, a two-step procedure is required. First, the $\hat{\bm{\beta}}_{\sst{M}}$ estimate is obtained  by fitting model~\eqref{eq:y} and implementing the corrections for $\beta_{\sst{INT}}$ and $\bm{\beta}_b$. Then, $\hat{\bm{\delta}}_{\sst{M}}$ is computed by fitting model~\eqref{eq:mygennow}, where $o(y,\bm{x}_{y};\hat{\bm{\beta}}_{\sst{M}})$ is included as an offset term. Notice that $\hat{\bm{\delta}}_{\sst{M}}$ is called a \emph{partial} M-estimator~\citep{StefanskiBoss2002}, since it requires the $\hat{\bm{\beta}}_{\sst{M}}$ estimate of the first step to be plugged-in in place of the unknown $\bm{\beta}$ vector~\citep{Randles1982}. In this case, knowledge of the adjusting terms $\log(k_1)$ and $\log(\bm{k}_{-1})$ is essential, since the estimated offset $o(y,\bm{x}_{y};\hat{\bm{\beta}}_{\sst{M}})$ contains $\hat{\beta}_{\sst{INT}}$ and $\hat{\bm{\beta}}_b$, rather than $\hat{\beta}_{\sst{INT}}^{\star}$ and $\hat{\bm{\beta}}_b^{\star}$, in its expression.

The variance-covariance matrix of the $\hat{\bm{\theta}}_{\sst{M}}$ estimator is given by
\[
V(\hat{\bm{\theta}}_{\sst{M}}) = \frac{1}{n} A(Y_i,\bm{\theta})^{-1}B(Y_i,\bm{\theta})\{A(Y_i,\bm{\theta})^{-1}\}^\top,
\]
where
\[
A(Y_i,\bm{\theta}) = E \Biggl\{ \frac{\partial \bm{\Psi}_i(\bm{\theta})}{\partial\bm{\theta}^\top} \Biggr\} \qquad B(Y_i,\bm{\theta}) = E  \Biggl\{ \bm{\Psi}_i(\bm{\theta}) \bm{\Psi}_i^\top(\bm{\theta}) \Biggr\}
\]
and $\bm{\Psi}_i(\bm{\theta})$ is the score random vector of a generic unit $i$. The finite-sample estimate of $V(\hat{\bm{\theta}}_{\sst{M}})$ can be computed as
\begin{equation}\label{eq:estvarcovM}
\hat{V}(\hat{\bm{\theta}}_{\sst{M}}) = \frac{1}{n} A(\bm{y},\hat{\bm{\theta}}_{\sst{M}})^{-1}B(\bm{y},\hat{\bm{\theta}}_{\sst{M}})\{A(\bm{y},\hat{\bm{\theta}}_{\sst{M}})^{-1}\}^\top,
\end{equation}
where $A(\bm{y},\bm{\theta})$ and $B(\bm{y},\bm{\theta})$ are the sample equivalent of $A(Y_i,\bm{\theta})$ and $B(Y_i,\bm{\theta})$~\citep{StefanskiBoss2002}. Their expressions are also reported in Appendix~\ref{app:mest}.


\subsection{Maximum Likelihood estimation}\label{subsec:mle}
Since we are dealing with a SCC sample, the likelihood of the observed data corresponds to the conditional density of the $(M,A,S,Z)$ random vector given $(Y,B)$ and $W=1$. As not only the $\bm{\beta}$ parameters of \eqref{eq:ypop} are the object of inference, but also the $\bm{\delta}$ parameters of \eqref{eq:mpop},  ML theory developed by Prentice and Pyke (1979)~\citep{PrenticePyke1979} should be extended. With reference to discrete choice models, Imbens (1992)~\citep{Imbens1992} provides more general results that can be applied to this context. Specifically, given $n$ independent sample units indexed by $i=1,\dots,n$, it follows from Imbens (1992) that ML estimation of $\bm{\theta}$ can be performed by maximizing

\begin{equation} \label{eq:Lk}
L(\bm{\theta}) = \prod_{i=1}^n P(Y_i=y_i,M_i=m_i\mid a_i,s_i,z_i,b_i,W_i=1).\\ 
\end{equation}

An additional factorization of~\eqref{eq:Lk} leads to
\[
L(\bm{\theta}) =\prod_{i=1}^n P(Y_i=y_i\mid a_i,s_i,z_i,b_i,W_i=1)P(M_i=m_i\mid a_i,s_i,z_i,b_i,y_i,W_i=1),
\]
so that the corresponding log-likelihood $\ell(\bm{\theta})=\log L(\bm{\theta})$ is equal to
\begin{equation}\label{eq:lk}
\begin{split}
\ell(\bm{\theta}) = \sum_{i=1}^n \ell_i(\bm{\theta}) &= \sum_{i=1}^n y_i \log p_{y^{\star}_{(m)},i} + (1-y_i)\log \{1-p_{y^{\star}_{(m)},i}\} \\
&\qquad\quad+ m_i \log p_{my,i} + (1-m_i) \log\{1- p_{my,i}\},
\end{split}
\end{equation}
where $p_{y^{\star}_{(m)},i}$ and $p_{my,i}$ are shorthands for $P(Y_i=y_i\mid a_i,s_i,z_i,b_i,W_i=1)$ and $P(M_i=m_i\mid a_i,s_i,z_i,b_i,y_i,W_i=1)$, respectively. While the latter is related to the parameter vector $\bm{\theta}$ via~\eqref{eq:mygennow}, the former involves marginalization over the binary mediator $M$. The corresponding model on the logistic scale can be written as
\begin{equation}\label{eq:mmargw}
\begin{split}
\textup{logit}\{P(Y=1\mid a,s,z,b,W=1)\} &= \textup{logit}\{P(Y=1\mid a,s,z,b,M=0, W=1)\} \\
&\quad+ g(\bm{x}_{y},\bm{x}_{m};\bm{\beta},\bm{\delta}), 
\end{split}
\end{equation}
where the first term is \eqref{eq:y} evaluated at $M=0$ while
\[
g(\bm{x}_{y},\bm{x}_{m};\bm{\beta},\bm{\delta}) = \textup{log}\frac{P(M=0\mid Y=0, a,s,z,b, W=1)} {P(M=0\mid Y=1, a,s,z,b, W=1)}
\]
is a correction term which depends on $\bm{x}_{y}$, $\bm{x}_{m}$ as well as on both parameter vectors. See Appendix~\ref{app:gterm} for the parametric expression of~\eqref{eq:mmargw}. 

Within this framework, the ML estimate of $\bm{\theta}$, $\hat{\bm{\theta}}_{\sst{ML}}$, can be obtained by maximizing the log-likelihood in~\eqref{eq:lk} via the usual iterative methods. Since these methods are likely to suffer from local maxima problems, it is advisable to set several starting points fluctuating around a sensible deterministic choice; see Section~\ref{subsec:res} for an account of the approach undertaken within our specific application. To further enhance the performance of optimization algorithms, it is typically useful to provide the expression of the log-likelihood gradient $\bm{s}(\bm{\theta})=\partial \ell(\bm{\theta})/\partial\bm{\theta}$, which is reported in the first part of Appendix~\ref{app:grad}.

In line with standard theory, the estimated variance-covariance matrix of $\hat{\bm{\theta}}_{\sst{ML}}$ is given by $\{-H(\hat{\bm{\theta}}_{\sst{ML}})\}^{-1}$, where $H(\hat{\bm{\theta}}_{\sst{ML}})=\{\partial \bm{s}(\bm{\theta})/\partial\bm{\theta}^\top\}_{\bm{\theta}=\hat{\bm{\theta}}_{\sst{ML}}}$ is the Hessian matrix computed at $\hat{\bm{\theta}}_{\sst{ML}}$. Alternatively, sandwich estimation~\citep{Royall1986,White1980} can be performed via
\[
\widehat{\textup{Cov}}(\hat{\bm{\theta}}_{\sst{ML}}) = \{-H(\hat{\bm{\theta}}_{\sst{ML}})\}^{-1}\bm{Q}(\hat{\bm{\theta}}_{\sst{ML}})\{-H(\hat{\bm{\theta}}_{\sst{ML}})\}^{-1},
\]
where $\bm{Q}(\hat{\bm{\theta}}_{\sst{ML}}) = \bigl\{\sum_{i=1}^n\bm{s}_i(\bm{\theta})\bm{s}_i(\bm{\theta})^\top \bigr\}_{\bm{\theta}=\hat{\bm{\theta}}_{\sst{ML}}}$ and $\bm{s}_i(\bm{\theta})$ is the individual contribution of the $i$-th unit to $\bm{s}(\bm{\theta})$.


\section{Case study}\label{sec:appl}
\subsection{The Listeriosis dataset}\label{subsec:data}
To illustrate the proposed approach, we reconsider the dataset analyzed by Preu{\ss}el et al.~\citep{Preusseletal2015}, where risk factors related to listeriosis in Germany are investigated. Listeriosis is an infection primarily contracted through the intake of food contaminated with \emph{Listeria monocytogenes} bacteria. It mainly affects older adults, individuals with weakened immune systems and pregnant women~\citep{Gouletetal2012}. Consequences might be quite severe, ranging from the development of life-threatening conditions for the fetus - in the case of pregnancy - to severe illness or death in the other cases.

The study by Preu{\ss}el et al.~\citep{Preusseletal2015} focusses on sporadic non-pregnancy associated listeriosis. Many food-related risk factors, like consumption of cold cooked sausages and pre-sliced cheese, are considered in combination with personal characteristics as well as other risk factors such as Reduced Immunocompetency (RI) or the intake of Gastric Acid Suppressors (GASs). The latter act as effect modifiers and can also be thought of as risk factors themselves when focus goes beyond food-related factors~\citep{Gouletetal2012,Hoetal1986,BavishiDupont2011,Mooketal2013}. In this application, we build upon this framework and extend it to a mediation setting in order to investigate whether - and to what extent - the additional vulnerability to listeriosis due to RI goes through the intake of GASs, which are commonly prescribed drugs for patients with RI~\citep{Preusseletal2015,Ahrensetal2012}. In line with the original approach of Preu{\ss}el et al.~\citep{Preusseletal2015}, the exposure (RI) is coded as a categorical variable with three mutually exclusive levels: 0=none, 1=RI due to immunocompromising diseases without immunosuppressive therapies (e.g., diabetes, autoimmune disorder), 2=RI due to immunosuppressive therapies (e.g., chemotherapy, radiation therapy). The mediator (GAS intake) is binary (0=no, 1=yes). RI and GAS intake are considered present if occurring within three months from listeriosis onset (cases) or interview (controls).

The dataset in Preu{\ss}el et al.~\citep{Preusseletal2015} contains 109 sporadic (i.e., not due to an outbreak) cases ascertained among over 40 men and non-pregnant women from all German Federal states (except Bremen, which accounts for around 0.8\% of the German population) in the time span from March 2012 until December 2013. In addition, 1982 controls were enrolled from the population of subjects with no history of listeriosis which are accessible by telephone in Germany. In the original study, controls were sampled to get an equal number of individuals in the three age classes 40-65 years, 66-75 years and 76+ years, according to the age distribution of cases in the 2004-2011 years. However, the age distribution of the 109 cases in the dataset differs from this one. While this fact might be due to a number of reasons, we argue that such a discrepancy does not allow one to treat the original dataset as a proper SCC one. Indeed, in an SCC design involving age as a stratifying factor, a fixed number of controls within the same age class for each case are sampled, resulting in exactly the same age distribution for cases and controls. Conversely, in an unconditional CC design the age distribution of the controls should be close to the age distribution of the population (differing by sampling error only). Since the dataset has neither of these features, we obviate by randomly selecting a fixed number of controls (8 times the number of cases) in a way such that the age distribution of the German population in 2012 (according to the  official webpage of the Federal Statistical Office of Germany) is matched, thereby generating an unconditional CC sample. The prevalence of sporadic non-pregnancy associated listeriosis is calculated assuming that 7 days is the average length of infection. The obtained rate is $1.70\times 10^{-7}$, which results in an adjusting correction term $\log(k_1)=13.511$.

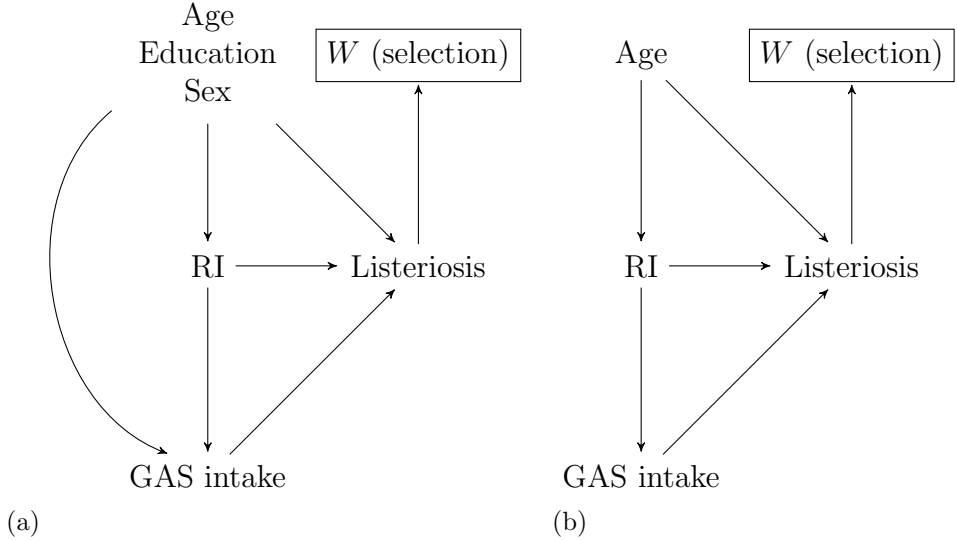
\begin{figure}[tb]
\centering{
\subfloat[][]{\label{fig:exfood}}
\begin{tikzpicture}[scale=0.5,auto,->,>=stealth',shorten >=1pt,node distance=2.8cm] 
\node (X) {RI};
\node (Y) [right of=X]{Listeriosis};
\node (M) [below of=X] {GAS intake};
\node[squarenode] (W) [above of=Y] {$W$ (selection)};
\node (B) [left of=W] {\begin{tabular}{c} Age \\ Education \\Sex \end{tabular}};
\draw[->] (X) --node {} (Y); 
\draw[->] (X) --node {} (M); 
\draw[->] (M) --node {} (Y); 
\draw[->] (Y) --node {} (W);
\draw[->] (B) --node {} (X); 
\draw[->] (B) --node {} (Y); 
\draw[->] (B) to [bend right=60] (M);
\end{tikzpicture}\quad
\subfloat[][]{\label{fig:exnofood}}
\begin{tikzpicture}[scale=0.5,auto,->,>=stealth',shorten >=1pt,node distance=2.8cm] 
\node (X) {RI};
\node (Y) [right of=X]{Listeriosis};
\node (M) [below of=X] {GAS intake};
\node (Z) [left of=W] {Age};
\node[squarenode] (W) [above of=Y] {$W$ (selection)};
\draw[->] (X) --node {} (Y); 
\draw[->] (X) --node {} (M); 
\draw[->] (M) --node {} (Y); 
\draw[->] (Y) --node {} (W);
\draw[->] (Z) --node {} (Y);
\draw[->] (Z) --node {} (X);
\end{tikzpicture}
}
\caption{DAGs representing the initially postulated (a) and final (b) scheme when investigating the association between RI and listeriosis through GAS.}
\label{fig:ex}
\end{figure}

In accordance with the data generating process postulated in Preu{\ss}el et al.~\citep{Preusseletal2015}, we initially include age, educational level and gender as covariates both for the outcome and mediation model (Figure~\ref{fig:exfood}). However, subsequent analyses lead to keeping only age (coded with the three classes introduced above) in the outcome model, and no covariates in the mediator model; see Figure~\ref{fig:exnofood}. With this regard, it is worth to underline that model selection is conducted separately for the two models, within an M-estimation framework. In detail, selection procedures for the mediator model are implemented conditionally on the estimated offset term, $o(y,\bm{x}_{y_0};\hat{\bm{\beta}}_{\sst{M}})$, resulting from the selected outcome model. Even at intermediate steps of the selection process, inference is based on the estimated variance-covariance matrix~\eqref{eq:estvarcovM} rather than on the naive standard errors returned by statistical software. 

\subsection{Results}\label{subsec:res}
Table~\ref{tab:regr} reports results for two pairs of logistic models fitted with M-estimation, ML and weighting. Standard errors are obtained with a sandwich approach not only for M- and ML estimation (see Section~\ref{sec:est}), but also for weighting. ML estimation is performed through direct maximization of the log-likelihood function by means of the \texttt{maxLik} function in \texttt{R}~\citep{HenningsenToomet2011}. Multiple starting values (10) are obtained by perturbing the $\hat{\bm{\theta}}_{\sst{M}}$ vector obtained from the M-estimates with a random normal deviation with null mean and standard deviation equal to 0.5. All attempts converge to the same solution. However, notice that unstable solutions might occur when starting values are determined totally at random.


\begin{table}[tb]
\centering
\begin{tabularx}{1.1\textwidth}{lcccccc}
\toprule
 & \multicolumn{2}{c}{\textbf{M-estimation}} & \multicolumn{2}{c}{\textbf{ML}} & \multicolumn{2}{c}{\textbf{Weighting}} \\
 \cmidrule(lr){2-3}\cmidrule(lr){4-5}\cmidrule(lr){6-7}
 & est. & s.e. & est. & s.e. & est. & s.e \\ 
 \midrule
\multicolumn{1}{c}{\textbf{Outcome model}} &&&&&&\\
Intercept & -17.089 & 0.297 & -17.104 & 0.298 & -17.250 & 0.299 \\ 
$\mathbb{I}\{\text{RI}=1\}$ & 0.871 & 0.564 & 0.901 & 0.555 & 0.798 & 0.566 \\ 
$\mathbb{I}\{\text{RI}=2\}$ & 2.696 & 0.393 & 2.729 & 0.391 & 2.668 & 0.398 \\ 
$\mathbb{I}\{\text{AGE}=66-75\}$ & 0.893 & 0.477 & 0.950 & 0.469 & 0.846 & 0.478 \\ 
$\mathbb{I}\{\text{AGE}=76+\}$ & 1.471 & 0.467 & 1.497 & 0.460 & 1.431 & 0.466 \\ 
$\mathbb{I}\{\text{RI}=1\}\cdot\mathbb{I}\{\text{AGE}=66-75\}$ & 1.239 & 0.752 & 1.145 & 0.738 & 1.492 & 0.793 \\ 
$\mathbb{I}\{\text{RI}=2\}\cdot\mathbb{I}\{\text{AGE}=66-75\}$ & -0.566 & 0.657 & -0.643 & 0.644 & -0.566 & 0.668 \\    
$\mathbb{I}\{\text{RI}=1\}\cdot\mathbb{I}\{\text{AGE}=76+\}$ & -0.441 & 0.780 & -0.482 & 0.775 & -0.248 & 0.813 \\ 
$\mathbb{I}\{\text{RI}=2\}\cdot\mathbb{I}\{\text{AGE}=76+\}$ & -1.537 & 0.673 & -1.624 & 0.675 & -1.358 & 0.703 \\ 
$\mathbb{I}\{\text{GAS}=1\}$ & 0.982 & 0.311 & 0.989 & 0.317 & 1.267 & 0.413 \\
\multicolumn{1}{c}{\textbf{Mediator model}} &&&&&&\\
Intercept & -3.276 & 0.219 & -3.276 & 0.218 & -3.153 & 0.208 \\ 
$\mathbb{I}\{\text{RI}=1\}$ & 0.845 & 0.330 & 0.843 & 0.335 & 0.497 & 0.421 \\ 
$\mathbb{I}\{\text{RI}=2\}$ & 0.841 & 0.347 & 0.838 & 0.351 & 0.576 & 0.472 \\  
\midrule
\multicolumn{1}{c}{\textbf{Outcome model}} &&&&&& \\
Intercept & -16.951 & 0.228 & -16.939 & 0.223 & -17.100 & 0.244 \\ 
$\mathbb{I}\{\text{RI}=1\}$ & 1.287 & 0.289 & 1.279 & 0.289 & 1.387 & 0.298 \\ 
$\mathbb{I}\{\text{RI}=2\}$ & 2.190 & 0.279 & 2.181 & 0.280 & 2.160 & 0.281 \\ 
$\mathbb{I}\{\text{AGE}=66-75\}$ & 1.014 & 0.263 & 1.016 & 0.261 & 0.972 & 0.283 \\ 
$\mathbb{I}\{\text{AGE}=76+\}$ & 0.711 & 0.307 & 0.689 & 0.309 & 0.684 & 0.312 \\ 
$\mathbb{I}\{\text{GAS}=1\}$ & 0.987 & 0.308 & 0.989 & 0.317 & 1.169 & 0.384 \\ 
\multicolumn{1}{c}{\textbf{Mediator model}} &&&&&& \\ 
Intercept & -3.276 & 0.219 & -3.276 & 0.218 & -3.153 & 0.208 \\ 
$\mathbb{I}\{\text{RI}=1\}$ & 0.844 & 0.332 & 0.843 & 0.335 & 0.497 & 0.421 \\
$\mathbb{I}\{\text{RI}=2\}$ & 0.839 & 0.344 & 0.838 & 0.351 & 0.576 & 0.472 \\
\bottomrule
\end{tabularx}
\caption{Estimates and standard errors (s.e.) of logistic models for the outcome and the mediator with (top) and without (bottom) AGE-RI interactions in the outcome model.\label{tab:regr}}
\end{table}


In the first pair of models (top part of Table~\ref{tab:regr}), the outcome equation contains interaction effects between age classes and RI, whereas in the second pair (bottom part of the table) it does not. Conversely, the mediator equation always includes RI effects only, meaning that its selection strategy is not influenced by whether or not interactions are included in the outcome model (parameter estimates are in practice not sensitive to this change, too). The choice between the outcome model with and without interactions is not straightforward. Indeed, only the coefficient for the $\mathbb{I}\{\textup{RI}=2\}\cdot\mathbb{I}\{\textup{AGE}=76+\}$ product is significant. However, an overall ANOVA test comparing them (within the M-estimation framework) provides evidence in favor of the extended one ($\chi^2$ deviance statistics equal to 11.706 with 4 degrees of freedom, p-value 0.020). Evidence is mixed also with regard to the Akaike and Bayesian Information Criteria, with the former favoring the interaction model (523.65 versus 527.35 for the no interaction model), and the latter favoring the no interaction model (556.11 versus 571.58 for the interaction model). For this reason, both pairs are kept for reference. In order to ease interpretation, the outcome regression coefficients of AGE and RI are also linearly combined to obtain contrasts, on the log odds-ratio scale, between any pair of levels. These are reported, together with their standard errors, in Table~\ref{tab:effects}, both for the interaction (top part) and no interaction (bottom part) setting. Clearly, in the former case the estimated contrasts of each factor are conditional on the level of the other.


\begin{table}[tb]
\centering
\begin{tabularx}{\textwidth}{lcccccc}
\toprule
 & \multicolumn{2}{c}{\textbf{M-estimation}} & \multicolumn{2}{c}{\textbf{ML}} & \multicolumn{2}{c}{\textbf{Weighting}} \\
 \cmidrule(lr){2-3}\cmidrule(lr){4-5}\cmidrule(lr){6-7}
 & est. & s.e. & est. & s.e. & est. & s.e \\ 
 \midrule
 \multicolumn{3}{c}{\textbf{Outcome model with interactions}} &&&& \\
& \multicolumn{6}{c}{$\text{AGE}=66-75$ vs $\text{AGE}=40-65$} \\
\cmidrule(lr){2-7}
$\text{RI}=0^\star$ & 0.893 & 0.477 & 0.950 & 0.469 & 0.846 & 0.478 \\
$\text{RI}=1$ & 2.132 & 0.577 & 2.095 & 0.569 & 2.338 & 0.609 \\
$\text{RI}=2$ & 0.327 & 0.450 & 0.307 & 0.441 & 0.280 & 0.470 \\
\cmidrule(lr){1-7}
& \multicolumn{6}{c}{$\text{AGE}=76+$ vs $\text{AGE}=40-65$} \\
\cmidrule(lr){2-7}
$\text{RI}=0^\star$ & 1.471  & 0.467 & 1.497  & 0.460 & 1.431 & 0.466 \\
$\text{RI}=1$ & 1.030  & 0.623 & 1.015  & 0.623 & 1.183 & 0.654 \\
$\text{RI}=2$ & -0.066 & 0.485 & -0.127 & 0.495 & 0.073 & 0.517 \\
\cmidrule(lr){1-7}
& \multicolumn{6}{c}{$\text{AGE}=76+$ vs $\text{AGE}=66-75$} \\
\cmidrule(lr){2-7}
$\text{RI}=0$ & 0.578  & 0.521 & 0.547  & 0.510 & 0.585  & 0.517 \\
$\text{RI}=1$ & -1.102 & 0.508 & -1.081 & 0.531 & -1.155 & 0.535 \\
$\text{RI}=2$ & -0.393 & 0.548 & -0.434 & 0.552 & -0.207 & 0.585 \\
\cmidrule(lr){1-7}
& \multicolumn{6}{c}{$\text{RI}=1$ vs $\text{RI}=0$} \\
\cmidrule(lr){2-7}
$\text{AGE} = 40-65^\star$ & 0.871 & 0.564 & 0.901 & 0.555 & 0.798 & 0.566 \\
$\text{AGE} = 66-75$ & 2.110 & 0.492 & 2.046 & 0.491 & 2.290 & 0.521 \\
$\text{AGE} = 76+$ & 0.430 & 0.536 & 0.418 & 0.541 & 0.550 & 0.562 \\
\cmidrule(lr){1-7}
& \multicolumn{6}{c}{$\text{RI}=2$ vs $\text{RI}=0$} \\
\cmidrule(lr){2-7}
$\text{AGE} = 40-65^\star$ & 2.696 & 0.393 & 2.729 & 0.391 & 2.668 & 0.398 \\
$\text{AGE} = 66-75$ & 2.131 & 0.524 & 2.086 & 0.513 & 2.102 & 0.529 \\
$\text{AGE} = 76+$ & 1.160 & 0.545 & 1.104 & 0.550 & 1.310 & 0.569 \\
\cmidrule(lr){1-7}
& \multicolumn{6}{c}{$\text{RI}=2$ vs $\text{RI}=1$} \\
\cmidrule(lr){2-7}
$\text{AGE} = 40-65$ & 1.825 & 0.545 & 1.828 & 0.537 & 1.870  & 0.547 \\
$\text{AGE} = 66-75$ & 0.021 & 0.485 & 0.040 & 0.485 & -0.188 & 0.523 \\
$\text{AGE} = 76+$ & 0.729 & 0.569 & 0.686 & 0.587 & 0.760  & 0.596 \\   
\bottomrule
\end{tabularx}
\caption{Estimated log odds-ratio contrasts between any pair of levels for outcome effects of $\text{AGE}$ and $\text{RI}$, both for the interaction (top) and no interaction (bottom) model. Stars denote contrasts related to regression coefficients, already in Table~\ref{tab:regr}.\label{tab:effects}}
\end{table}

\begin{table}[tb]
\centering
\begin{tabularx}{\textwidth}{lcccccc}
\toprule
 & \multicolumn{2}{c}{\textbf{M-estimation}} & \multicolumn{2}{c}{\textbf{ML}} & \multicolumn{2}{c}{\textbf{Weighting}} \\
 \cmidrule(lr){2-3}\cmidrule(lr){4-5}\cmidrule(lr){6-7}
 & est. & s.e. & est. & s.e. & est. & s.e \\ 
 \midrule
\multicolumn{3}{c}{\textbf{Outcome model without interactions}} &&&& \\
& \multicolumn{6}{c}{$\text{AGE}$} \\
\cmidrule(lr){2-7}
$66-75$ vs $40-65^\star$ & 1.014 & 0.263 & 1.016 & 0.261 & 0.972 & 0.283 \\ 
$76+$ vs $40-65^\star$ & 0.711 & 0.307 & 0.689 & 0.309 & 0.684 & 0.312 \\ 
$76+$ vs $66-75$ & -0.303 & 0.105 & -0.327 & 0.107 & -0.287 & 0.110 \\
\cmidrule(lr){1-7}
& \multicolumn{6}{c}{$\text{RI}$} \\
\cmidrule(lr){2-7}
1 vs 0$^\star$ & 1.287 & 0.289 & 1.279 & 0.289 & 1.387 & 0.298 \\ 
2 vs 0$^\star$ & 2.190 & 0.279 & 2.181 & 0.280 & 2.160 & 0.281 \\
2 vs 1 & 0.903 & 0.285 & 0.901 & 0.287 & 0.773 & 0.301 \\
\bottomrule
\end{tabularx}
\caption*{Table~\ref{tab:effects} continues.}
\end{table}


Tables~\ref{tab:regr} and~\ref{tab:effects} show that ML and M-estimation provide very similar results for both model pairs. In particular, the estimated coefficient for the effect of GAS intake on listeriosis development is always positive (very close to 0.99, p-value 0.002). This confirms that GAS intake remains positively associated with listeriosis, even when controlling for RI~\citep{Preusseletal2015}. Similar conclusions can be drawn for RI effects due to both diseases (RI=1) and therapies (RI=2) on GAS intake, with all coefficients being around 0.84 (p-values lower than 0.002). In light of these results, we can conclude that there is no evidence of a difference between diseases and therapies with respect to their effect on GAS intake.

With regard to effects of age on the outcome, the interaction model shows an increasing pattern for the no RI (RI=0) group, although the shift from the $40-65$ to the $66-75$ age class is barely significant for M-estimation (p-value 0.061), and that from the $66-75$ to the $76+$ class is not significant (p-values greater than 0.25). For RI=1, a rather relevant positive difference is estimated for the contrast involving the second and the first age class. Conversely, the estimated difference between the third and the first class is smaller (and barely significant, p-values around 0.010). As a consequence, the risk for listeriosis in the second class is higher than that of the third one, with the difference being significant (p-value 0.030 for M-estimation and 0.042 for ML). For the RI=2 group, no significant differences among age classes are present. In the no interaction model, a unique pattern is estimated for all RI groups which is similar to the one described for the RI=1 group in the model with interactions, although with reduced magnitudes.

As for RI effects on the outcome, in the no interaction model we notice a significantly increased vulnerability to listeriosis when moving from RI=0 to both RI=1 and RI=2 (p-values lower than 0.001). The difference between the latter levels is significant, too, with immunosuppressive therapies further enhancing the risk for listeriosis with respect to immunocompromising diseases (p-values 0.002). Effect magnitudes are in line with the findings in the original study; see Preu{\ss}el et al.~\citep{Preusseletal2015}. In the model with interactions, such a pattern of effect directions is preserved within each age class. However, the shifts from RI=0 to RI=1 and from RI=1 to RI=2 are significant only for the second and the first class, respectively (p-values lower than 0.002). Overall, we can conclude that: i) only therapies have a significant effect in the $40-65$ class, ii) differences between therapies and diseases are essentially null in the $66-75$ class, and iii) RI effects considerably lessen in the $76+$ class. 

In terms of effect direction and statistical significance, results from the weighting approach are not extremely dissimilar to those previously reported. However, some noteworthy discrepancies emerge which concern GAS intake effects in the outcome model and RI effects in the mediator model. For the former, we observe a sensible increase in the coefficients with respect to M- and ML estimation. Moreover, it is important to remark that these estimates are somewhat less robust to model specification, with two rather distant values in the models with and without AGE-RI interactions (1.267 and 1.169, respectively). As for RI effects on GAS intake, the comparison with the other two approaches yields smaller coefficients, with no significant differences between any pair of levels.

\subsubsection{Causal interpretation of effects}\label{subsubsec:causal}
Causal mediation analysis requires a number of assumptions to hold~\citep{VdWVansteelandt2010}. These include \emph{consistency} and \emph{composition}~\citep{VdW2009,VdWVan2009}, as well as more context-specific assumptions concerning the absence of unobserved confounding for the following relationships: i) exposure-outcome, ii) mediator-outcome, iii) exposure-mediator. Also, iv) no mediator-outcome confounding should be generated by variables causally affected by the treatment. In this case, many effects of interest are not identified unless strong parametric assumptions are made, even when these variables are measured~\citep{VdWVansteelandt2010,Pearl2014}. While consistency and composition are technical statements typically tenable in many applied settings, assumptions i)-iv) warrant some discussion about the introduced covariate sets. As mentioned in Section~\ref{subsec:data}, for our application we directly rely on the data generating process of Preu{\ss}el et al.~\citep{Preusseletal2015}, from which it seems reasonable to assume that the included covariates are sufficient for the assumptions i)-iii) to hold. Also, none of these variables is causally affected by the exposure, so that assumption iv) seems realistic, too.

When the outcome is binary, VanderWeele and coauthors~\citep{VdWVansteelandt2010,ValeriVdW2013} introduce the notion of Natural Direct Effect (NDE) and Natural Indirect Effect (NIE) on the odds-ratio scale. In particular, when also the mediator is binary, Valeri and VanderWeele~\citep{ValeriVdW2013} derive parametric expressions under the rare outcome assumption. Though this is also our context, we rely on the formulae in Doretti et al. (2022)~\citep{Dorettietal2022}, that extend them to settings where the exposure (and the mediator) interact with the covariates (also outside the rare outcome case). When the binary mediator does not interact with other variables, it can be shown that exposure effects in the outcome model (that is, RI effects in Table~\ref{tab:effects}) are good approximations of log odds-ratio NDEs. We have also computed NIEs on the same scale: for our data, these depend very little on the covariate patterns (with differences at the sixth digit), even in the model with interactions.

With regard to M- and ML estimation, the NIEs related to the $\text{RI}=2$ vs $\text{RI}=1$ contrast are always very close to 0 and not significant. Clearly, this is due to the almost null difference between the effects of these two exposure levels in the mediator model (see the corresponding regression coefficients in Table~\ref{tab:regr}). Conversely, when the reference level of $\text{RI}$ is 0, NIEs lie around 0.068, with Delta-method~\citep{Oehlert1992} standard errors close to 0.042 (p-value 0.105). An exception is represented by the $\text{RI}=2$ vs $\text{RI}=0$ contrast in the setting with interactions, where the NIE has the same magnitude but the standard error is 0.037 (p-value 0.066). As for weighting, the NDE and NIE values are slightly different, reflecting the fact that the estimated regression coefficient differ from those of the other approaches; see the related discussion in Section~\ref{subsec:res}. However, the interpretation of results is essentially the same. In Table~\ref{tab:prop}, we report the proportion of RI effect mediated by GAS intake (for the 1 vs 0 and the 2 vs 0 contrasts) on the log odds-ratio scale. On such a scale, total causal effects are given by the sum of NDEs and NIEs. This proportion is small but non-negligible, since it ranges from 2.4\% to 14\% across RI contrasts and age classes. Nevertheless, it is worth to remark that the significance of estimated NIEs is weak, and that total causal effects are non-significant whenever the corresponding NDEs also are.


\begin{table}[tb]
\centering
\begin{tabularx}{0.9\textwidth}{lYYYYYY}
\toprule
& \multicolumn{2}{c}{\textbf{M-estimation}} & \multicolumn{2}{c}{\textbf{ML}} & \multicolumn{2}{c}{\textbf{Weighting}} \\
 \cmidrule(lr){2-3}\cmidrule(lr){4-5}\cmidrule(lr){6-7}
RI contrast & 1 vs 0 & 2 vs 0 & 1 vs 0 & 2 vs 0 & 1 vs 0 & 2 vs 0 \\
\midrule
& \multicolumn{6}{c}{\textbf{Outcome model with interactions}} \\
 \cmidrule(lr){2-7}
$\text{AGE}=40-65$ & 0.072 & 0.024 & 0.070 & 0.024 & 0.065 &  0.024 \\ 
$\text{AGE}=66-75$ & 0.031 & 0.031 & 0.032 & 0.031& 0.024 & 0.031 \\ 
$\text{AGE}=76+$ & 0.136 & 0.055  & 0.140 & 0.058 & 0.091 & 0.048\\ 
\midrule
 & \multicolumn{6}{c}{\textbf{Outcome model without interactions}} \\
 \cmidrule(lr){2-7}
$\text{AGE}=\text{All}$ & 0.050 & 0.030 & 0.051 & 0.030 & 0.034 & 0.022  \\ 
\bottomrule
\end{tabularx}
\caption{Proportion of RI effect mediated by GAS intake\label{tab:prop}}
\end{table}


\section{Simulation study}\label{sec:sims}
In this section, we present a simulation study conducted to investigate the finite-sample properties of the M- and ML estimators introduced in Section~\ref{sec:est}. The performance of these estimators is compared to that of the weighting estimator, implemented for both the outcome and mediator model. In Section~\ref{subsec:simdes}, the structure of the study is described, while results are reported in Section~\ref{subsec:simres}.

\subsection{Design}\label{subsec:simdes}

\begin{figure}[tb]
 \includegraphics[scale=0.2]{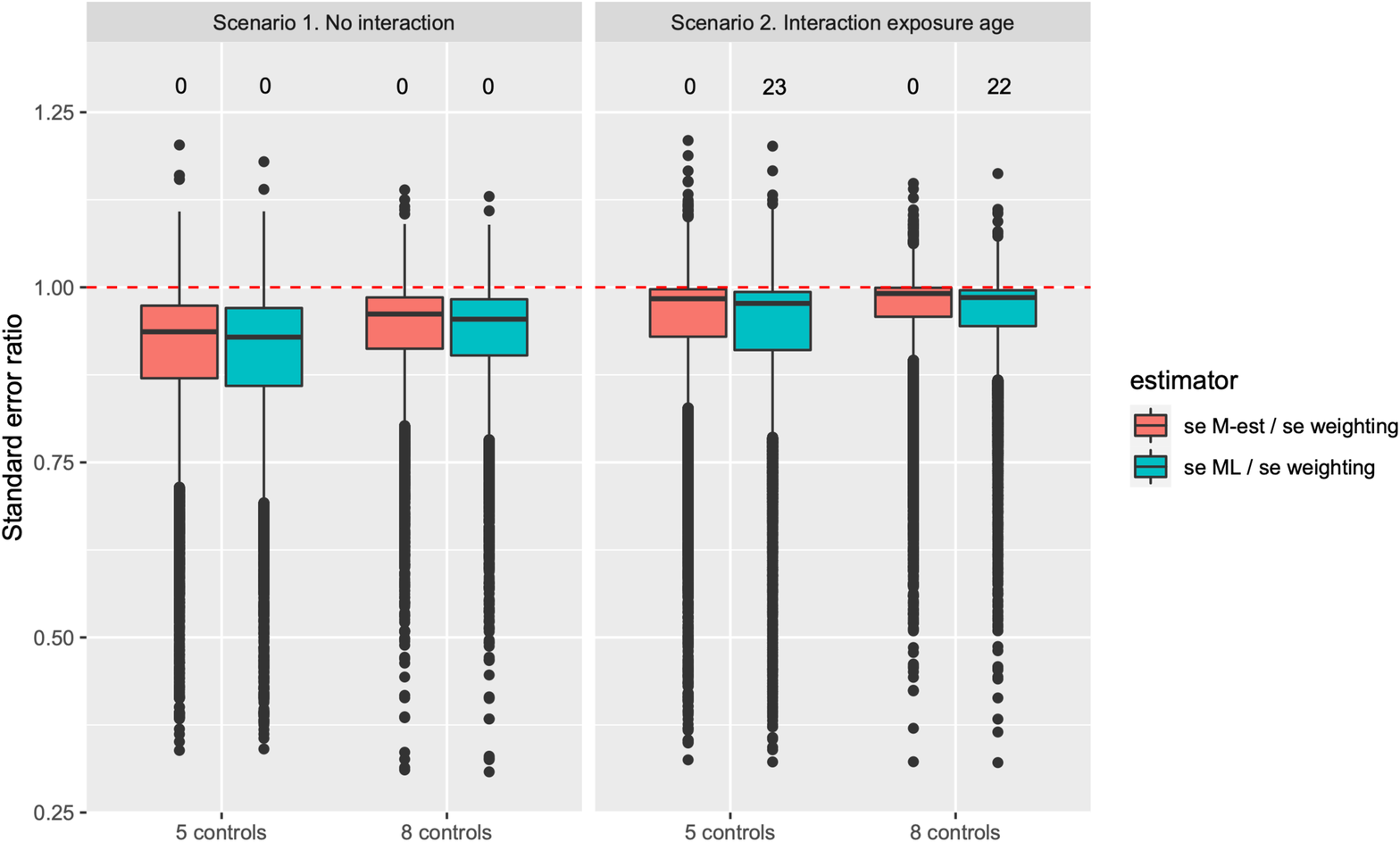}
 \caption{Box plots of the quotient between the estimated standard errors of the ML / M-estimator and the weighting estimator for the regression parameters $\bm{\beta}$ and $\bm{\delta}$ (i.e. 1000 replications of 9 or 13 parameters). Note that we have changed the limits of the y-axis in order to make the figure more clear, the number of observations above the window is printed above each box.}
 \label{se:fig}
 \end{figure}
 
The simulation study mimics the two scenarios of the case study in Section \ref{sec:appl} by using the same distribution of the covariates and taking the parameter values for~\eqref{eq:ypop} and~\eqref{eq:mpop} quite close to those estimated from that dataset. The only exception concerns the outcome model intercept, which is raised in order to get a higher expected prevalence. In this way, at least 100 cases for each simulated population are obtained. We generate 1000 populations of 30 million individuals. More specifically, the covariates are simulated randomly with $P(\mbox{RI}=0)=0.73$, $P(\mbox{RI}=1)=0.16$, $P(\mbox{RI}=2)=0.11$, $P(\text{AGE}= 40-65)=0.66$, $P(\text{AGE}=66-75)=0.19$, $P(\text{AGE}=76+)=0.15$. In both scenarios, the mediator (GAS intake) is generated by $M \sim \text{Be}(\text{expit}(\bm{x}_m^\top\bm{\delta}))$, with $\bm{x}_m^\top= (1,\mathbb{I}\{\text{RI}=1\},\mathbb{I}\{\text{RI}=2\})$ and $\bm{\delta}=(-3.3, 0.8, 0.8)^\top$. As for the outcome (listeriosis onset), we have $Y \sim \text{Be}(\text{expit}(\bm{x}_y^\top\bm{\beta}))$. In Scenario 1 (no interaction model) $\bm{x}_y$ is as in the bottom part of Table~\ref{tab:regr} and $\bm{\beta}=(-12.4, 1.3, 2.2, 1.0, 0.7, 1.0)^\top$, whereas in Scenario 2 (interaction model) $\bm{x}_y$ is as in the top part of Table~\ref{tab:regr} and $\bm{\beta}=(-13.1, 0.9, 2.7, 0.9, 1.5, 1.2, -0.6, -0.5, -1.6, 1)^\top$. For each generated population, we randomly select 100 cases and 500 or 800 controls. Notice that every time the marginal population prevalence, $\pi$, can be computed exactly.

\subsection{Results}\label{subsec:simres}
The regression parameter estimates $\hat{\bm{\beta}}$ and $\hat{\bm{\delta}}$ of the ML and M-estimator are quite similar and show that these estimators perform slightly better than the weighting estimator; see Table~\ref{tab:5cov} for the setting with 5 controls per case (with 8 controls similar results are obtained). Importantly, these findings suggest that the estimators derived within our analytical framework generally overcome the weighting approach not only for the outcome model parameters (see Section~\ref{sec:est}), but also for the mediator model parameters. The ML estimator is the most efficient when the maximization is stable. For 21 out of 1000 repetitions of Scenario 2, the maximization was however unstable, resulting in one or more extreme estimated standard errors (up to 3.5 million times the size of the corresponding standard error of the weighting estimator). The M-estimator is more robust compared to the ML estimator when it comes to such extreme results and performs slightly better compared to the weighting estimator when it comes to precision; see Figure \ref{se:fig}. The bias is very low for almost all parameters, with two exceptions. First, the $\mathbb{I}\{\text{RI}=1\}\cdot\mathbb{I}\{\text{AGE}=76+\}$ coefficient (scenario 2); the empirical coverage is however as desired (at least $95\%$). Second, the intercept from the outcome model of both scenarios and sample sizes is biased for the weighting estimator, often resulting in slightly too low empirical coverage (see the values in Table~\ref{tab:5cov} for the setting with 5 controls per case, while results for the larger sample size are not reported).

\begin{table}[tb]
\begin{tabular}{rrrrrrrrrrrrr}
\toprule
 & \multicolumn{3}{c}{\textbf{Bias}} & \multicolumn{3}{c}{\textbf{MC SD}} & \multicolumn{3}{c}{\textbf{RMSE}}& \multicolumn{3}{c}{\textbf{EC}} \\
 \cmidrule(lr){2-4}\cmidrule(lr){5-7}\cmidrule(lr){8-10}\cmidrule(lr){11-13}
 & M-est & ML &W&  M-est & ML & W & M-est & ML & W & M-est & ML & W \\ 
  \midrule
 \multicolumn{3}{c}{\textbf{Scenario 1}} &&&&\\
 \multicolumn{3}{l}{\textbf{Outcome model}} &&&&\\
 & -0.02 & -0.02 & -0.27 & 0.22 & 0.22 & 0.24 & 0.22 & 0.22 & 0.36 & 0.97 & 0.96 & 0.88 \\ 
 & 0.00 & 0.00 & 0.00 & 0.31 & 0.31 & 0.32 & 0.31 & 0.31 & 0.32 & 0.95 & 0.95 & 0.95 \\
 & 0.02 & 0.02 & 0.05 & 0.30 & 0.30 & 0.32 & 0.30 & 0.30 & 0.32 & 0.94 & 0.94 & 0.94 \\ 
 & -0.01 & -0.01 & 0.01 & 0.29 & 0.28 & 0.31 & 0.29 & 0.28 & 0.31 & 0.95 & 0.95 & 0.96 \\ 
 & -0.01 & -0.01 & 0.01 & 0.34 & 0.33 & 0.36 & 0.34 & 0.33 & 0.36 & 0.95 & 0.94 & 0.95 \\ 
& 0.01 & 0.00 & 0.07 & 0.42 & 0.40 & 0.49 & 0.42 & 0.40 & 0.50 & 0.95 & 0.95 & 0.93 \\ 
 \multicolumn{3}{l}{\textbf{Mediator model}} &&&&\\
 & -0.05 & -0.05 & -0.04 & 0.29 & 0.29 & 0.30 & 0.29 & 0.29 & 0.30 & 0.95 & 0.95 & 0.96 \\ 
 & -0.01 & -0.01 & -0.04 & 0.45 & 0.45 & 0.63 & 0.45 & 0.45 & 0.63 & 0.96 & 0.96 & 0.97 \\ 
 & -0.04 & -0.03 & -0.26 & 0.45 & 0.44 & 1.36 & 0.45 & 0.44 & 1.39 & 0.96 & 0.96 & 0.96 \\ 
   \multicolumn{3}{c}{\textbf{Scenario 2}} &&&&\\
   \multicolumn{3}{l}{\textbf{Outcome model}} &&&&\\
& -0.05 & -0.05 & -0.29 & 0.31 & 0.31 & 0.32 & 0.32 & 0.31 & 0.43 & 0.96 & 0.96 & 0.93 \\ 
& -0.05 & -0.05 & -0.06 & 0.79 & 0.78 & 0.73 & 0.79 & 0.79 & 0.73 & 0.96 & 0.96 & 0.96 \\ 
 & 0.06 & 0.06 & 0.07 & 0.42 & 0.41 & 0.43 & 0.42 & 0.42 & 0.43 & 0.95 & 0.95 & 0.95 \\ 
& 0.02 & 0.02 & 0.02 & 0.53 & 0.53 & 0.53 & 0.53 & 0.53 & 0.53 & 0.95 & 0.95 & 0.95 \\ 
& 0.00 & -0.00 & 0.00 & 0.47 & 0.46 & 0.47 & 0.47 & 0.46 & 0.47 & 0.96 & 0.96 & 0.96 \\ 
& 0.07 & 0.07 & 0.09 & 1.00 & 0.99 & 0.96 & 1.00 & 0.99 & 0.97 & 0.95 & 0.95 & 0.94 \\ 
 & -0.01 & -0.01 & -0.01 & 0.76 & 0.75 & 0.79 & 0.76 & 0.75 & 0.79 & 0.96 & 0.96 & 0.95 \\ 
& -0.18 & -0.18 & -0.13 & 2.13 & 2.11 & 1.88 & 2.14 & 2.12 & 1.89 & 0.97 & 0.97 & 0.95 \\ 
 & -0.06 & -0.06 & -0.03 & 1.49 & 1.48 & 1.69 & 1.49 & 1.48 & 1.69 & 0.95 & 0.95 & 0.95 \\ 
& 0.01 & -0.00 & 0.10 & 0.43 & 0.40 & 0.53 & 0.43 & 0.40 & 0.54 & 0.95 & 0.95 & 0.94 \\ 
   \multicolumn{3}{l}{\textbf{Mediator model}} &&&&\\
& -0.05 & -0.04 & -0.04 & 0.29 & 0.29 & 0.30 & 0.29 & 0.29 & 0.31 & 0.95 & 0.95 & 0.96 \\ 
 & -0.02 & -0.02 & -0.05 & 0.45 & 0.45 & 0.56 & 0.45 & 0.45 & 0.56 & 0.95 & 0.95 & 0.96 \\ 
 & -0.02 & -0.01 & -0.21 & 0.50 & 0.49 & 1.38 & 0.50 & 0.49 & 1.40 & 0.94 & 0.94 & 0.96 \\ 
   \hline
\end{tabular}
        \caption{Empirical bias, Monte Carlo Standard Deviation (MC SD), Root Mean Square Error (RMSE) and Empirical Coverage (EC) of $95\%$ confidence intervals constructed using estimated standard errors (sandwich estimators). We use 5 controls per case, the weighting estimator is abbreviated to W. Within each scenario, the parameter sequence is as in Table~\ref{tab:regr}.}
        \label{tab:5cov}
\end{table}


\section{Discussion}\label{sec:concl}
We consider Stratified Case-Control (SCC) designs defined on a certain variable (primary outcome) and we address the problem of modeling another binary variable (secondary outcome) from the resulting dataset. In detail, we assume that a logistic regression model for the secondary outcome holds in the target population, and we recover the parameters of such a model via a simple offset adjustment. Like in other frameworks, knowledge of the (conditional) prevalence(s) of the primary outcome in the population is required.

Our analytical solution allows one to perform mediation analysis from SCC data in settings where both the outcome and the mediator are binary. As shown in Section~\ref{sec:est}, the derived offset correction leads to M- and Maximum Likelihood (ML) estimation of the joint (i.e., outcome and mediator) model parameter vector, provided that the intercept and/or the coefficients of background variables in the outcome model are suitably adjusted. These estimation methods are opposed to weighting, where such adjustments are not needed. With this respect, a noticeable finding of the simulation study in Section~\ref{sec:sims} is that the efficiency gain of M- and ML estimation compared to weighting is not limited to the outcome model vector, but it extends to the mediator model parameters. Although we have explored a single scenario (close to that of the application described in Section~\ref{sec:appl}), we reasonably believe that a similar tendency would be observed for alternative parameter configurations, too.

The proposed approach is exemplified via the analysis of a CC dataset gathered within a German study on listeriosis conducted by Preu{\ss}el et al.~\citep{Preusseletal2015}. Specifically, we take a novel perspective and focus on evaluating whether the effect of Reduced Immunocompetency (RI) on listeriosis development is mediated by the intake of Gastric Acid Suppressors (GASs). We try to answer this question also in causal terms by decomposing the total causal effect of RI on listeriosis into the Natural Direct Effect (NDE) and the Natural Indirect Effect (NIE), though we acknowledge that such a terminology might be somewhat questionable from a strict causal mediation standpoint. This is because RI and GAS intake do not indeed \emph{cause} listeriosis; the ingestion of food contaminated with \emph{Listeria monocytogenes} bacteria (predominantly) or other unknown vehicles (to a lesser extent) do. Nevertheless, since infection means are not of primary interest, the disentanglement of the RI effect on listeriosis in causal terms appears sensible, due to the clear additional vulnerability to listeriosis for people with RI as well as for people taking GASs, and to the fact that GAS intake is a direct consequence of RI. Results show that the NIE of RI is almost negligible.

It is important to stress that any conclusion from the reported analyses has to be drawn cautiously, for several reasons. First, in the original study by Preu{\ss}el et al.~\citep{Preusseletal2015}, 732 patients with listeriosis were potentially eligible as cases. Of these, only 109 entered the study due to participation refusal, decease, inability to answer/be contacted and other unknown reasons. Also, like in many telephone surveys recruited controls were found to systematically differ from the target population with respect to socio-economic status. Although included and not included cases present not too dissimilar distributions for many variables, and socio-economic status is essentially accounted for via the (initial) inclusion of educational level as its proxy~\citep{Preusseletal2015}, we cannot exclude that some degree of non-informative missingness~\citep{Molenberghs2014handbook} might affect the results. 

Furthermore, as discussed in Section~\ref{subsubsec:causal}, though the typical assumptions of causal mediation analysis i)-iv) seem to be tenable in our application, it is worth to remark that these requirements about confounding might be more difficult to justify in other contexts. To deal with violations of i)-iii), sensitivity analyses~\citep{Lindmarketal2018} as well as methods embedded in an instrumental variable framework~\citep{didelez2010assumptions,mattei2011augmented} have been introduced which could be possibly adapted to the present setting. As for iv), alternative causal estimands identifiable also in the presence of exposure-induced mediator-outcome confounding (named \emph{interventional effects}) were proposed~\citep{VdWetal2014,VansteelandtDaniel2017}.

It is well-known that in SCC designs the number of strata can in principle increase until the limit situation of containing a single case (sometimes termed \emph{exact matching}). In these settings, traditional estimation might be problematic, with conditional likelihood methods typically preferable~\citep{Gailetal1981,Breslow1996}.  Extension to deal with these cases can be of interest.


\appendix


\section{Proof of (6) and~(7)}\label{app:der45}
By standard probability results it is possible to express the left-hand side of~\eqref{eq:mygennow} as
\begin{equation}\label{eq:proof45}
\begin{split}
\log\frac{P(M=1\mid a,b,s,z,y)}{P(M=0\mid a,b,s,z,y)} &= \log\frac{P(Y=y\mid a,b,s,z,M=1)}{P(Y=y\mid a,b,s,z,M=0)} + \log\frac{P(M=1\mid a,b,s,z)}{P(M=0\mid a,b,s,z)} \\
&= \log\frac{P(Y=y\mid a,b,z,M=1)}{P(Y=y\mid a,b,z,M=0)} + \log\frac{P(M=1\mid a,b,s)}{P(M=0\mid a,b,s)} \\
&= \log\frac{P(Y=y\mid a,b,z,M=1)}{P(Y=y\mid a,b,z,M=0)} +\bm{x}_m^\top\bm{\delta}.\\
\end{split}
\end{equation}

For the outcome model~\eqref{eq:ypop}, the partition $\bm{x}_y=(\bm{x}_{y_0}^\top,\bm{x}_{y_1}^\top)^\top$ is introduced, where $\bm{x}_{y_0}$ ($\bm{x}_{y_1}$) denotes the sub-vector of covariate values not involving (involving) $M$. The coefficient vector partition $\bm{\beta}=(\bm{\beta}_0^\top,\bm{\beta}_1^\top)^\top$ is defined accordingly. We assume our model to be hierachical, and therefore in the most general setting $\bm{x}_{y_1}=m\cdot\bm{x}_{y_0}$. However, in many applications $\bm{\beta}_1$ and $\bm{x}_{y_1}$ are likely to be smaller-dimensional than $\bm{\beta}_0$ and $\bm{x}_{y_0}$, with some elements of the latter vectors not having a counterpart in the former. In these cases, the additional vector $\tilde{\bm{\beta}}_1$ has to be defined by suitably expanding $\bm{\beta}_1$ with zeros, so that the sum of conformable vectors $\bm{\beta}_{\sst{+}}=\bm{\beta}_0+\tilde{\bm{\beta}}_1$ can be computed. It then follows that
\[
\begin{split}
\textup{logit}\{P(Y=1\mid a,b,z,M=0)\} &= \bm{x}_{y_0}^\top\bm{\beta}_0 \\
\textup{logit}\{P(Y=1\mid a,b,z,M=1)\} &= \bm{x}_{y_0}^\top\bm{\beta}_{\sst{+}}.
\end{split}
\]
Then,
\begin{equation}\label{eq:proof451}
\begin{split}
o(y,\bm{x}_{y};\bm{\beta})=\log\frac{P(Y=y\mid a,b,z,M=1)}{P(Y=y\mid a,b,z,M=0)} &= \log\frac{\exp\{y\cdot\bm{x}^\top_{y_0}\bm{\beta}_{\sst{+}}\}}{1+\exp\{\bm{x}^\top_{y_0}\bm{\beta}_{\sst{+}}\}} - \log\frac{\exp\{y\cdot\bm{x}^\top_{y_0}\bm{\beta}_0\}}{1+\exp\{\bm{x}^\top_{y_0}\bm{\beta}_0\}}  \\
&= y(\bm{x}^\top_{y_0}\bm{\beta}_{\sst{+}} - \bm{x}^\top_{y_0}\bm{\beta}_0) - \log\frac{1+\exp\{\bm{x}^\top_{y_0}\bm{\beta}_{\sst{+}} \}}{1+\exp\{\bm{x}^\top_{y_0}\bm{\beta}_0 \}}. \\
\end{split}
\end{equation}


\section{Details of M-estimation}\label{app:mest}
In order to obtain the expressions of the observed score function vectors $\bm{s}_y(\cdot)$ and $\bm{s}_m(\cdot)$, it is worth to rely on matrix notation. Specifically, we index sample units by $i=1,\dots,n$ and introduce the sample vectors $\bm{m}=(m_1,\dots,m_n)^\top$ and $\bm{y}=(y_1,\dots,y_n)^\top$, as well as the sample design matrices
\[
\bm{X}_m = \begin{pmatrix} \bm{x}_{m,1}^\top \\
					   \vdots  \\
					   \bm{x}_{m,n}^\top \end{pmatrix},
\qquad\qquad\qquad
\bm{X}_y = \begin{pmatrix} \bm{X}_{y_0} & \bm{X}_{y_1} \end{pmatrix} = \begin{pmatrix}
															\bm{x}_{y_0,1}^\top & \bm{x}_{y_1,1}^\top \\
															\vdots & \vdots \\
															\bm{x}_{y_0,n}^\top & \bm{x}_{y_1,n}^\top
														\end{pmatrix}.
\]
We also denote by $d_{\delta}$, $d_{\beta}$, $d_{\beta_0}$ and $d_{\beta_1}$ the number of columns of $\bm{X}_m$, $\bm{X}_y$, $\bm{X}_{y_0}$ and $\bm{X}_{y_1}$, respectively.

Letting $\textup{expit}(\cdot)=\exp(\cdot)/\{1+\exp(\cdot)\}$, for a given value $\bm{\theta}=(\bm{\beta}^\top,\bm{\delta}^\top)^\top$ we have
\[
\begin{split}
\bm{s}_y(\bm{\theta}) &= \bm{X}_y^\top(\bm{y}-\bm{p}_{y^{\star}}) \;\;\;\;\;\qquad \bm{p}_{y^{\star}} = \textup{expit}(\bm{\eta}_{y^{\star}}) \;\;\;\;\qquad \bm{\eta}_{y^{\star}} = \bm{X}_y\bm{\beta}^{\star}\\
\bm{s}_m(\bm{\theta}) &= \bm{X}_m^\top(\bm{m}-\bm{p}_{my}) \qquad\;\;\;\; \bm{p}_{my} = \textup{expit}(\bm{\eta}_{my}) \qquad\;\;\; \bm{\eta}_{my} = \bm{X}_m\bm{\delta} + \bm{o},
\end{split}
\]
where $\bm{\beta}^{\star}$ is like in Section~\ref{sec:ident} and $\bm{o}$ is a column vector collecting the offset terms $o(y_i,\bm{x}_{y_0,i};\bm{\beta})$ of every sample unit $i=1,\dots,n$. Notice that the vector $\bm{p}_{my}=(p_{my,1},\dots,p_{my,n})^\top$ contains the $p_{my,i}$ probabilities ($i=1,\dots,n$) already introduced in Section~\ref{subsec:mle}, while $\bm{p}_{y^\star}=(p_{y^{\star},1},\dots,p_{y^{\star},n})^\top$ differs from $\bm{p}_{y^{\star}_{(m)}}=(p_{y^{\star}_{(m)},1},\dots,p_{y^{\star}_{(m)},n})$, which is the vector collecting the $p_{y^{\star}_{(m)},i}$ probabilities (also used in~\ref{subsec:mle}). A compact form for $\bm{\psi}(\bm{\theta})=(\bm{s}_y(\bm{\theta})^\top,\bm{s}_m(\bm{\theta})^\top)^\top$ is given by
\[
\bm{\psi}(\bm{\theta}) = \bm{X}_j^\top\bm{f}, 
\]
where
\[
\bm{X}_j = \begin{pmatrix} \bm{X}_y & \bm{0}_{n\times d_{\delta}} \\ \bm{0}_{n\times d_{\beta}} & \bm{X}_m \end{pmatrix} \qquad \bm{f} =  \begin{pmatrix} \bm{y}-\bm{p}_{y^\star} \\ \bm{m}-\bm{p}_{my} \end{pmatrix}.
\]

The formula for the $A(\bm{y},\bm{\theta})$ matrix is
\[
A(\bm{y},\bm{\theta}) = \frac{1}{n}\begin{pmatrix} \bm{H}_{y\beta} & \bm{H}_{y\delta} \\ \bm{H}_{m\beta} & \bm{H}_{m\delta}  \end{pmatrix},
\]
where each generic sub-matrix is defined as $\bm{H}_{q\alpha}=\partial\bm{s}_q/\partial\bm{\alpha}^\top$ and
\[
\begin{split}
\bm{H}_{y\beta} &= -\bm{X}_y^\top\textup{diag}\{\bm{p}_{y^\star}\cdot(1-\bm{p}_{y^\star})\}\bm{X}_y \qquad\;\;\; \bm{H}_{y\delta} =  \bm{0}_{d_{\beta}\times d_{\delta}} \\
\bm{H}_{m\beta} &= -\bm{X}_m^\top\textup{diag}\{\bm{p}_{my}\cdot(1-\bm{p}_{my})\}\bm{D} \qquad\;\;\; \bm{H}_{m\delta} = -\bm{X}_m^\top\textup{diag}\{\bm{p}_{my}\cdot(1-\bm{p}_{my})\}\bm{X}_m.\\
\end{split}
\]
In the above, the $\bm{D}$ matrix is given by
\[
\bm{D} = \begin{pmatrix} \bm{0}_{n\times d_{\beta_0}} & \bar{\bm{X}}_{y_0}\cdot \bm{y} \end{pmatrix} + \begin{pmatrix} \bm{X}_{y_0}\cdot\bm{v}_0 & \bar{\bm{X}}_{y_0}\cdot \bm{v}_1 \end{pmatrix},
\]
where $\bar{\bm{X}}_{y_0}$ is the matrix obtained by extracting the columns of $\bm{X}_{y_0}$ corresponding to the non-null elements of $\tilde{\bm{\beta}}_1$ while $\bm{v}^0$ and $\bm{v}^1$ are two column vectors of length $n$, whose $i$-th element is given by
\[
v_{0,i} =  \frac{\exp(\bm{x}_{y_0,i}^\top\bm{\beta}_0) - \exp(\bm{x}^\top_{y_0,i}\bm{\beta}_{\sst{+}})}{\{ 1 +\exp(\bm{x}^\top_{y_0,i}\bm{\beta}_0)\} \{ 1 + \exp(\bm{x}^\top_{y_0,i}\bm{\beta}_{\sst{+}}) \}} \qquad\qquad v_{1,i} = - \frac{\exp(\bm{x}^\top_{y_0,i}\bm{\beta}_{\sst{+}})}{1+\exp(\bm{x}^\top_{y_0,i}\bm{\beta}_{\sst{+}})}.
\]
Notice that in the formulae above $\cdot$ denotes element-wise product, be it between column vectors or between a matrix and a column vector. In the latter case, the result is a conformable matrix where every column is element-wise multiplied by the vector.

Finally, $B(\bm{y},\bm{\theta})$ is given by
\[
B(\bm{y},\bm{\theta}) = \frac{1}{n} \sum_{i=1}^n \bm{\psi}_i(\bm{\theta}) \bm{\psi}_i^\top(\bm{\theta}),
\]
with $\bm{\psi}_i(\bm{\theta})$ being the contribution of unit $i$ to the observed score vector.


\section{Parametric expression of (11)} \label{app:gterm}
The parametric expression of $g(\bm{x}_{y_0},\bm{x}_{m};\bm{\beta},\bm{\delta})$ can be derived from \eqref{eq:proof45} and \eqref{eq:proof451}. After some simplifications, it follows that
\begin{equation}\label{Appendice3eq1}
g(\bm{x}_{y_0},\bm{x}_{m};\bm{\beta},\bm{\delta}) = \log\frac{\exp(\bm{x}_{y_0}^\top\tilde{\bm{\beta}}_1)\exp(\bm{x}_m^\top\bm{\delta})\{1+\exp(\bm{x}_{y_0}^\top\bm{\beta}_0) \} + \{ 1+\exp(\bm{x}_{y_0}^\top \bm{\beta}_{\sst{+}}) \}}{\exp(\bm{x}_m^\top\bm{\delta})\{1+\exp(\bm{x}^\top_y \bm{\beta}_0) \} + \{ 1+\exp(\bm{x}^\top_y \bm{\beta}_{\sst{+}}) \}},
\end{equation}
where the elements in the right-hand side are defined as in Appendix \ref{app:der45}. We then have
\[
\begin{split}
\textup{logit}\{P(Y=1\mid a,s,z,b,W=1)\}&=\bm{x}_{y_0}^\top\bm{\beta}_0^\star + g(\bm{x}_{y_0},\bm{x}_{m};\bm{\beta},\bm{\delta}),\\
\end{split}
\]
with $\bm{\beta}_0^\star$ being the modification of $\bm{\beta}_0$ accounting for the correction for $\beta_{\sst{INT}}$ and $\bm{\beta}_b$. 


\section{Gradient function for ML estimation}\label{app:grad}
The log-likelihood gradient $\bm{s}(\bm{\theta})$ can be conveniently expressed in matrix form. To this end, we maintain notation as in the body of the paper and in previous appendices, and we introduce the matrices
\[
\bm{A}_g = \begin{pmatrix} \bm{A}_{gy_0} \\ \bm{A}_{gy_1} \\ \bm{A}_{gm} \end{pmatrix} \qquad \bm{B}_g = \begin{pmatrix} \bm{B}_{gy_0} \\ \bm{B}_{gy_1} \\ \bm{B}_{gm} \end{pmatrix} \qquad
\bm{A}_m = \begin{pmatrix} \bm{A}_{my_0} \\ \bm{A}_{my_1} \\ \bm{A}_{mm} \end{pmatrix} \qquad \bm{B}_m = \begin{pmatrix} \bm{B}_{my_0} \\ \bm{B}_{my_1} \\ \bm{B}_{mm}\end{pmatrix},
\]
where
\[
\begin{split}
\bm{A}_{gy_0} &= \bm{1}_{d_{\beta_0}\times n} \qquad\qquad \bm{B}_{gy_0} =  \bm{G}_{\beta_{\sst{INT}}} \otimes \bm{1}_{d_{\beta_0}} \qquad\qquad \bm{A}_{my_0} = \bm{0}_{d_{\beta_0}\times n} \qquad\qquad \bm{B}_{my_0} = \bm{v}_0^\top \otimes \bm{1}_{d_{\beta_0}} \\
\bm{A}_{gy_1} &= \bm{0}_{d_{\beta_1}\times n} \qquad\qquad \bm{B}_{gy_1} =  \bm{G}_{\beta_m} \otimes \bm{1}_{d_{\beta_1}} \;\;\qquad\qquad \bm{A}_{my_1} = \bm{y}^\top \otimes \bm{1}_{d_{\beta_1}} \qquad\; \bm{B}_{my_1} = \bm{v}_1^\top \otimes \bm{1}_{d_{\beta_1}} \\
\bm{A}_{gm} &= \bm{0}_{d_{\delta}\times n} \qquad\qquad\;\; \bm{B}_{gm} = \bm{G}_{\delta_{\sst{INT}}} \otimes \bm{1}_{d_{\delta}} \qquad\qquad\;\;  \bm{A}_{mm} = \bm{1}_{d_{\delta}\times n} \qquad\qquad\; \bm{B}_{mm} = \bm{0}_{d_{\delta}\times n}. \\
\end{split}
\]
In the above, $\otimes$ denotes Kronecker product, while 
\[
\begin{split}
\bm{G}_{\beta_{\sst{INT}}} &= \begin{pmatrix}
\frac{\partial}{\partial \beta_{\sst{INT}}}g(\bm{x}_{y_0,1},\bm{x}_{m,1};\bm{\beta},\bm{\delta}) & \cdots & \frac{\partial}{\partial \beta_{\sst{INT}}}g(\bm{x}_{y_0,n},\bm{x}_{m,n};\bm{\beta},\bm{\delta})
\end{pmatrix} \\
\bm{G}_{\beta_m} &= \begin{pmatrix}
\frac{\partial}{\partial \beta_m}g(\bm{x}_{y_0,1},\bm{x}_{m,1};\bm{\beta},\bm{\delta}) & \cdots & \frac{\partial}{\partial \beta_m}g(\bm{x}_{y_0,n},\bm{x}_{m,n};\bm{\beta},\bm{\delta})
				\end{pmatrix} \\
\bm{G}_{\delta_{\sst{INT}}} &= \begin{pmatrix} \frac{\partial}{\partial \delta_{\sst{INT}}}g(\bm{x}_{y_0,1},\bm{x}_{m,1};\bm{\beta},\bm{\delta}) & \cdots & \frac{\partial}{\partial \delta_{\sst{INT}}}g(\bm{x}_{y_0,n},\bm{x}_{m,n};\bm{\beta},\bm{\delta})
				\end{pmatrix}
\end{split}
\]
are three row vectors collecting the derivatives of the $g(\bm{x}_{y_0,i},\bm{x}_{m,i};\bm{\beta},\bm{\delta})$ terms ($i=1,\dots,n$) with respect to $\beta_{\sst{INT}}$, $\beta_m$ and $\delta_{\sst{INT}}$, where the latter two are the coefficient of $M$ in the outcome model and the intercept of the mediation model, respectively. These derivatives can be computed by noticing that the argument of the logarithm can be written as $(q_1q_2q_3+q_4)/(q_2q_3+q_4)$, with $q_1=\exp(\bm{x}_{y_0,i}^\top\tilde{\bm{\beta}}_1)$, $q_2=\exp(\bm{x}_{m,i}^\top\bm{\delta})$, $q_3=1+\exp(\bm{x}_{y_0,i}^\top\bm{\beta}_0)$ and $q_4=1+\exp(\bm{x}_{y_0,i}^\top \bm{\beta}_{\sst{+}})$. Consequently, we have
\[
\begin{split}
\frac{\partial g}{\partial\beta_{\sst{INT}}} &= \frac{\{q_1q_2(q_3-1)+q_4-1\}(q_2q_3+q_4) - (q_1q_2q_3+q_4)\{q_2(q_3-1)+q_4-1\}}{\exp(g)(q_2q_3+q_4)^2}  \\
\frac{\partial g}{\partial\beta_m} &= \frac{(q_1q_2q_3+q_4-1)(q_2q_3+q_4)-(q_1q_2q_3+q_4)(q_4-1)}{\exp(g)(q_2q_3+q_4)^2}  \\
\frac{\partial g}{\partial\delta_{\sst{INT}}} &= \frac{(q_1q_2q_3)(q_2q_3+q_4) - (q_1q_2q_3+q_4)(q_2q_3)}{\exp(g)(q_2q_3+q_4)^2}.  \\
\end{split}
\]

Finally, letting
\[
\bm{C} = \begin{pmatrix} \bm{X}_{y_0} & \bar{\bm{X}}_{y_0} & \bm{X}_m \end{pmatrix}^\top,
\]
the log-likelihood gradient is given by
\begin{equation}\label{eq:lkgrad}
s(\bm{\theta}) = \{(\bm{A}_g+\bm{B}_g)\cdot \bm{C}\}\tilde{\bm{y}} + \{(\bm{A}_m+\bm{B}_m)\cdot \bm{C}\}\tilde{\bm{m}},
\end{equation}
where $\tilde{\bm{y}} = \bm{y} - \bm{p}_{y^{\star}_{(m)}}$, $\tilde{\bm{m}} = \bm{m} - \bm{p}_{my}$ and $\bm{p}_{my}$ and $\bm{p}_{y^{\star}_{(m)}}$ are like in Appendix~\ref{app:mest}.

\section*{Acknowledgement}
We acknowledge the financial support from Swedish Research Council for health working life and welfare [2019-01064].

\bibliographystyle{plain}

\end{document}